\documentclass{aa}
\usepackage{graphics}
\usepackage{amssymb}
\begin{document}
\thesaurus{ 08 % A&A Section 6: Form. struct. and evolut. of stars
           (08.02.2;  % binaries: eclipsing
            08.03.5;  % binaries: spectroscopic
            08.13.2;  % Stars: mass-loss
            08.09.2;  % Stars: individual: HR 6902
            13.21.5)} % Ultraviolet: stars

\title{The warm circumstellar envelope and wind of the G9 IIb star HR~6902}

\author{T. Kirsch  
\and R. Baade 
\and D. Reimers}

\institute{Hamburger Sternwarte, Universit\"at Hamburg, 
Gojenbergsweg~112, D-21029~Hamburg, Germany}

\offprints {T. Kirsch (tkirsch@hs.uni-hamburg.de) }

\date{Received  ; accepted , }

\maketitle

%\titlerunning{The circumstellar envelope of HR 6902}{}
%\authorrunning{T. Kirsch et al.}

\begin{abstract}
{\it IUE\/} observations of the eclipsing binary system HR 6902 obtained at 
various epochs spread over four years indicate the presence of warm 
circumstellar material enveloping the G9~IIb primary. The spectra show 
\ion{Si}{iv} and \ion{C}{iv} absorption up to a distance of 3.3 giant radii (${\rm R_g}$).
Line ratio diagnostics yields an electron temperature of $\rm \sim 78\,000\; K$ 
which appears to be constant over the observed height range.

Applying a least square fit absorption line analysis we derive column densities 
as a function of height. We find that the inner envelope ($< 3\;{\rm R_g}$) of 
the bright giant is consistent with a hydrostatic density distribution. The 
derived line broadening velocity of $\sim 70\; {\rm km\, s^{-1}}$ is sufficient 
to provide turbulent pressure support for the required scale height. However, 
an improved agreement with observations over the whole height regime including 
the emission line region is obtained with an outflow model. We demonstrate that 
the common $\beta$~power-law as well as a $P \propto \rho$ wind yield appropriate fit 
models. Adopting a continuous mass outflow we obtain a mass-loss rate of 
$\dot{M}=0.8 - 3.4 \times 10^{-11}$ $\rm M_{\odot}\,{yr^{-1}}$ depending on the 
particular wind model.

The emission lines observed during total eclipse are attributed mostly to resonance 
scattering of B star photons in the extended envelope of the giant. By means of a 
multi-dimensional line formation study we show that the global envelope properties 
are consistent with the wind models derived from the absorption line analysis. 
We argue that future high resolution UV spectroscopy will resolve the large-scale 
velocity structure of the circumstellar shell. As an illustration we present 
theoretical \ion{Si}{iv} and \ion{C}{iv} emission profiles showing model-dependent 
line shifts and asymmetries. 

\end{abstract}

\section{Introduction} 
\label{sec1} 

The spectroscopic binary system HR~6902 (G9~IIb + B8-9~V) belongs to the category of 
$\zeta$~Aur stars, where a hot companion (generally a main sequence B~star) moves 
through the extended envelope of an evolved late-type star. The specific binary 
geometry provides an exceptional opportunity to probe the spatial dependence of 
the physical properties of the outer atmosphere and wind of the primary. Observations 
of $\zeta$~Aur systems and related objects with the {\it International Ultraviolet 
Explorer\/} ({\it IUE\/}) and more recently with the {\it Hubble Space Telescope\/} 
({\it HST\/}) take advantage of the ultraviolet spectral region. At wavelengths 
below $\sim 3000\;\mathrm{\AA}$ the hot secondary dominates the observed flux 
allowing a separation of the spectral information. In the UV the observed spectrum 
is purely that of the companion with superimposed circumstellar and interstellar lines. 
Further information on the binary technique can be found in several reviews 
(e.g., \cite{rei89}; \cite{gui90}; \cite{ahm93}; \cite{har96}; \cite{baa98}).

Ground based spectroscopic as well as UV and X-ray observations from space of stars 
in the cool half of the HR diagram have shown that the extended atmospheres of 
giants and supergiants undergo a dramatic physical change across a region in 
the HR diagram. To the right of this transition we observe cool and massive winds 
plus extended cool chromospheres, while to its left there is much hotter circumstellar 
matter. This transition is accompanied by different spectral indicators defining 
distinct dividing lines. The leftmost is the "X-ray dividing line". All stars to 
its left show X-ray emission, probably arising from extended corona-like plasma 
(e.g., Haisch, Schmitt, \& Rosso 1991, 1992). In the ultraviolet the same stars also 
show \ion{C}{iv} and \ion{Si}{iv} and often \ion{N}{v} emission (\cite{lin79}).
Additional dividing lines are defined by asymmetries in the \ion{Ca}{ii} and 
\ion{Mg}{ii} emission lines (e.g., \cite{ste80a}, 1980b). The strongest and 
most reliable evidence for the existence of a cool, massive wind is the appearance 
of circumstellar absorption features in the \ion{Ca}{ii} line (\cite{rei77}). A
theoretical discussion of the dividing line concept based on hydromagnetic waves 
can be found in Rosner (1993) and Rosner et al.\ (1995).

However, this simple picture is complicated by the existence of the so-called 
hybrid stars. Hartmann, Dupree \& Raymond (1980) and Reimers (1982) found that
some bright K giants and G supergiants show blue-shifted \ion{Mg}{ii} 
(and partly \ion{Ca}{ii}) absorption and emission lines like \ion{C}{iv} and \ion{Si}{iv}
signifying temperatures of at least $10^5\;{\rm K}$. Hybrid stars are also X-ray 
emitters (\cite {rei96}). The outflow properties of these stars are still very 
uncertain and it is controversial whether hybrids show cold ($\sim 10^4\;{\rm K}$) 
or possibly warm ($\sim 10^5\; {\rm K}$) winds.

Previous {\it IUE\/} observations of HR~6902 suggest the existence of warm 
circumstellar material that may represent the transitional physics of hybrid 
star atmospheres. Reimers, Baade, \& Schr\"oder (1990a) and Reimers et al.\
(1990b) have presented a series of {\it IUE\/} high resolution spectra ($R \sim 10^4$)
taken a few days before and after eclipse. \ion{Si}{iv} and \ion{C}{iv} column densities 
indicate a height-independent electron temperature of $T_{\rm {e}} \approx 70\,000\;
{\rm K}$. As discussed by Reimers et al.\ (1990b) the density distribution is 
compatible with a low mass-loss wind.

In the present paper we report the results of a reanalysis of {\it IUE\/} spectra
of HR~6902. Compared to the first study (\cite{rei90a}, 1990b) we utilize a 
substantially expanded observations database. The paper is organized as follows:
In Sect.~\ref{sec2} we present the observations and basic stellar parameters
used in this paper. Sect.~\ref{sec3} summarizes the methodical principles
of our absorption analysis technique and the application to the HR~6902 
spectra. In Sect.~\ref{sec4} we outline the multi-dimensional radiative transfer 
procedure used to infer global properties of the circumstellar shell.

\section{Observations and basic stellar data}
\label{sec2}

While nearly all observations have been conducted by ourselves in several 
{\it IUE\/} programs, the data used here are taken from the {\it IUE\/} Final 
Archive. We have compiled a sequence of large aperture spectra obtained between 
1987 and 1990. Full details of the {\it NEWSIPS\/} data reduction have been given 
by Garhart et al.\ (1997), and will not be repeated here. It is, however, worth 
reiterating that the new processing techniques generally greatly enhance the quality of 
{\it IUE\/} spectra. The increase in signal-to-noise improves the detection limit 
of faint absorption features significantly. Furthermore, the new noise handling 
allows a reliable error estimate for each point in the flux spectrum. The archival 
{\it IUE\/} spectra exploited for this paper are listed in Table~\ref{tbl-1}.

The analysis presented by Reimers et al.\ (1990b) is based on {\it IUE\/} 
observations of the 1988 and 1989 eclipses, covering a range of impact parameters 
$p$ (projected distance between the stellar components) from $\sim 1.1 - 2.0$ 
giant radii. Our new study also includes the high resolution spectra of the 1990 
eclipse showing \ion{Si}{iv} and \ion{C}{iv} absorption up to $p = 3.3\;{\rm R_g}$.
These additional impact parameters allow us to directly examine the wind
acceleration region. One additional observation obtained 1987 yields an upper limit 
to the column densities at $p = 5.3\;{\rm R_g}$.

\begin{table}
\caption[]{{\it IUE\/} observations of HR 6902. All exposures were taken through the large aperture}
\begin{flushleft}
\begin{tabular}{lcccc}
\hline
\noalign{\vspace {0.1cm}}
 Date &SWP &Exp.\ time &$p$  & Phase\\
      &  No.   & (min)& $\mathrm{(R_g)}$  &     \\
\noalign{\vspace {0.1cm}}
\hline
\noalign{\vspace {0.1cm}}
1987    &      &    &      &\\
Jul. 11 &31325 &140 &5.26 &0.9227\\
1988    &      &    &      &\\
Aug. 25 &34131 &140 &1.15 &0.9887\\
Aug. 25 &34135 &110 &1.10 &0.9898\\
Sep. 03 &34177 &130 &1.19 &0.0122\\
Sep. 04 &34182 &125 &1.33 &0.0147\\
Sep. 05 &34187 &100 &1.48 &0.0175\\
1989    &      &    &     &\\
Jun. 30 &36589 &140 &9.12 &0.7926\\
Sep. 26 &37191 &180 &1.68 &0.0209\\
Sep. 27 &37195 &190 &1.84 &0.0235\\
Sep. 28 &37200 &194 &2.00 &0.0261\\
1990    &      &     &     &\\
Sep. 20 &39668 &195 &3.32 &0.9534\\
Sep. 20 &39669 &195 &3.29 &0.9538\\
Sep. 23 &39699 &195 &2.81 &0.9612\\
Sep. 23 &39700 &180 &2.79 &0.9616\\
Sep. 26 &39713 &177 &2.28 &0.9694\\
Sep. 28 &39721 &170 &1.98 &0.9742\\
Oct. 08 &39795$^{\rm a}$ &410 &0.84 &0.9997\\
\noalign{\vspace {0.1cm}}
\hline
\end{tabular}
\end{flushleft}
$^{\mathrm{a}}$ Low-dispersion spectrum
\label{tbl-1}
\end{table}

\begin{figure}
\resizebox{\hsize}{!}{\includegraphics{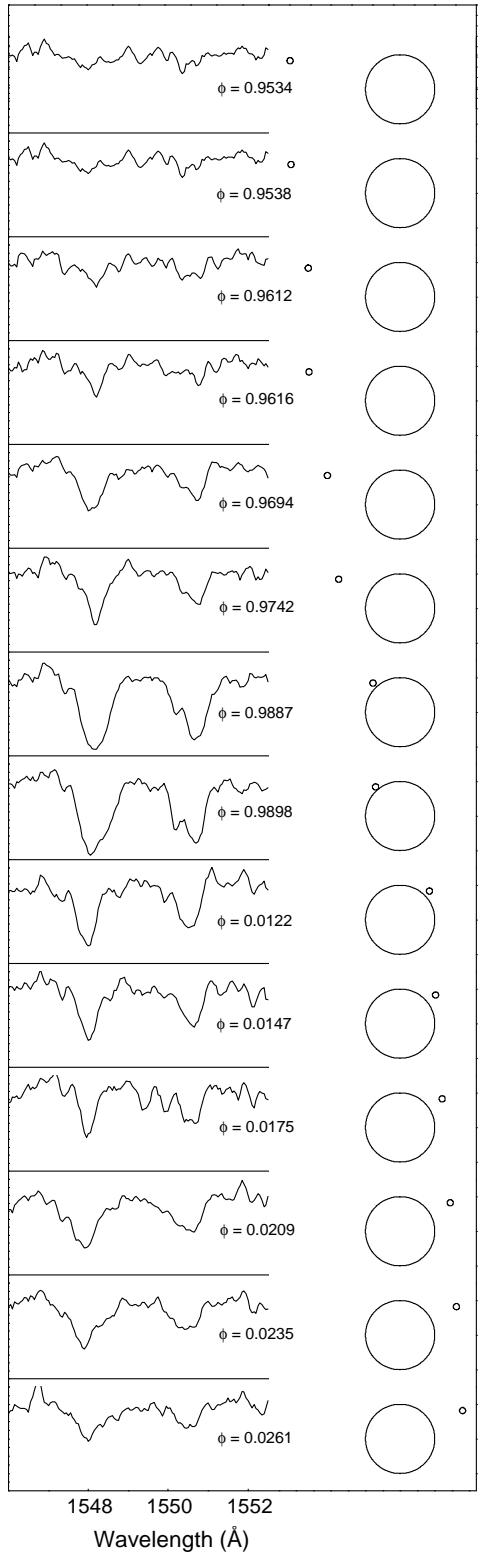}}
\caption[]{Series of {\it IUE\/} high resolution spectra of 
\ion{C}{iv} UV~mult.~1 and the corresponding positions of the B star behind 
the primary as seen from Earth. The spectra are smoothed with a 
Savitzky-Golay filter}
\label{fig1}
\end{figure}

A direct analysis of the circumstellar \ion{Si}{iv} and \ion{C}{iv} lines is not 
feasible owing to the considerable photospheric background absorption. We extract 
the circumstellar contribution by dividing all high resolution spectra by the 
out-of-eclipse exposure SWP 36589. This observation serves as suitable reference 
spectrum since the large impact parameter of $p~= 9.1\;{\rm R_g}$ implies a pure 
B~star spectrum. Indeed, the final wind model presented in Sect.~\ref{sec3} confirms that there 
should be no detectable circumstellar absorption. As an example we present in 
Fig.~\ref{fig1} the phase variation of the \ion{C}{iv} absorption line. The observed line 
strength depends strongly on the relative position of the B star and decreases with 
increasing impact parameter. Though circumstellar \ion{C}{ii} absorption at 1335~\AA\
is clearly detectable, the poor S/N ratio of the {\it IUE\/} spectra does not allow an adequate 
absorption line reconstruction. In the case of nearly saturated photospheric 
absorption lines the dividing procedure leads to spurious line profiles with 
artificial emission features. Unfortunately, the \ion{C}{iv} and \ion{Si}{iv} doublets are the only
circumstellar absorption lines that can be used for a quantitative analysis.

\begin{table}
\caption[]{Orbital elements and stellar parameters of HR 6902 [gathered from Griffin et al.\ (1995); 
the distance is obtained from the {\it Hipparcos\/} Main Catalog]}
\begin{flushleft}
\begin{tabular}{lc}
\hline
\noalign{\vspace {0.1cm}}
 Parameter & Value\\
\noalign{\vspace {0.1cm}}
\hline
\noalign{\vspace {0.1cm}}
Period (days) &385.0\\
Eccentricity&0.311\\
Inclination (deg)&87\\
Longitude of periastron (deg)&146 \\
Semimajor axis (cm)& $\mathrm{\,2.94 \times 10^{13}}$ \\
Periastron passage (JD)&2\,447\,306.33\\
Conjunction (JD)&2\,447\,788.18\\
Spectral type (primary)&G9 IIb\\
Spectral type (secondary)&B8-9 V\\
$R_\mathrm{G}/\mathrm{{R_{\odot}}}$&33.0\\
$R_\mathrm{B}/\mathrm{{R_{\odot}}}$&3.0\\
$M_\mathrm{G}/\mathrm{{M_{\odot}}}$&3.86\\
$M_\mathrm{B}/\mathrm{{M_{\odot}}}$&2.95\\
$T_\mathrm{eff,G}$ (K)&4900\\
$T_\mathrm{eff,B}$ (K)&11\,600\\
$M_\mathrm{V,G}$ (mag)& $-1.81$\\
$M_\mathrm{V,B}$(mag)&$-0.03$\\
$E_\mathrm{B-V}$(mag)&$-0.18$\\
Distance (pc)&$247 \pm 40$\\
\noalign{\vspace {0.1cm}}
\hline
\end{tabular}
\end{flushleft}
\label{tbl-2}
\end{table}

\begin{figure}
\resizebox{\hsize}{!}{\includegraphics{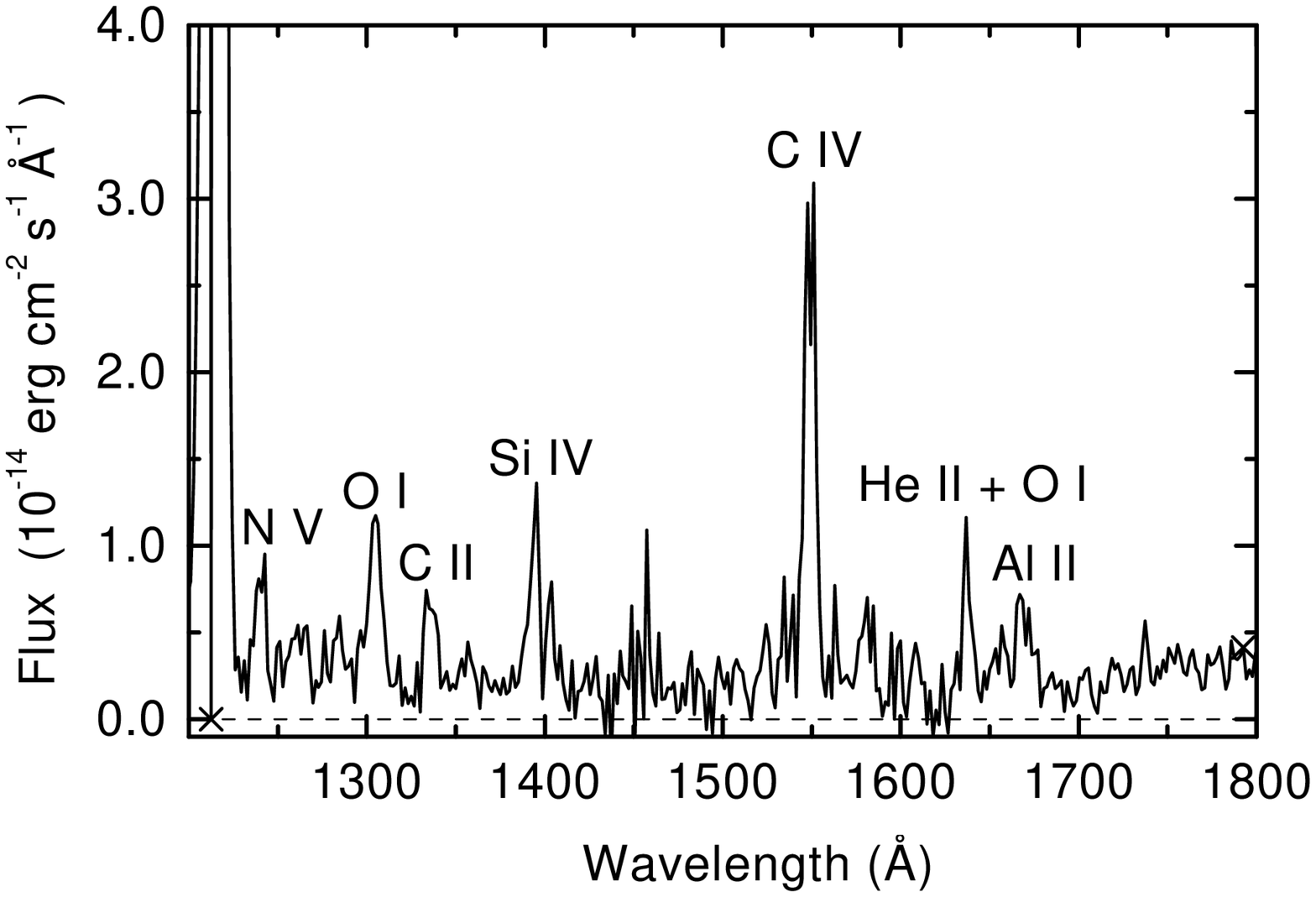}}
\caption[]{{\it IUE\/} low resolution spectrum SWP 39795 
obtained at phase $\phi = 0.9997$ showing the 
\ion{C}{iv} and \ion{Si}{iv} resonance doublets and a number of additional
emission lines}
\label{fig2}
\end{figure}

Great significance must be attached to the emission line spectrum visible during 
total eclipse of the B star. The 1988 total eclipse observation (SWP 34143, $\phi = 0.998$) 
was part of the first HR~6902 program and has been reported by Reimers et al.\ (1990a). 
The deep $410^{\rm m}$ exposure SWP 39795 (Fig.~\ref{fig2}) obtained on 1990 October 8 
provides a significantly improved signal-to-noise ratio and will be subject of 
this paper. We note that the large difference in quality does not allow a 
comparative analysis in order to search for long-term variations. While in 
"normal" $\zeta$~Aur systems hundreds of scattering lines (e.g., \ion{Fe}{ii} 
and \ion{Si}{ii}) are seen, only a few are observed in the spectrum of HR~6902. 
A comparison with stars of similar spectral type suggests that the emission line 
spectrum cannot be due to the G~star alone (\cite{rei90a}). Instead, the observed 
\ion{C}{iv} and \ion{Si}{iv} doublets are formed primarily by resonance scattering of 
B~star photons in the circumstellar envelope.    

\begin{figure}
\resizebox{\hsize}{!}{\includegraphics{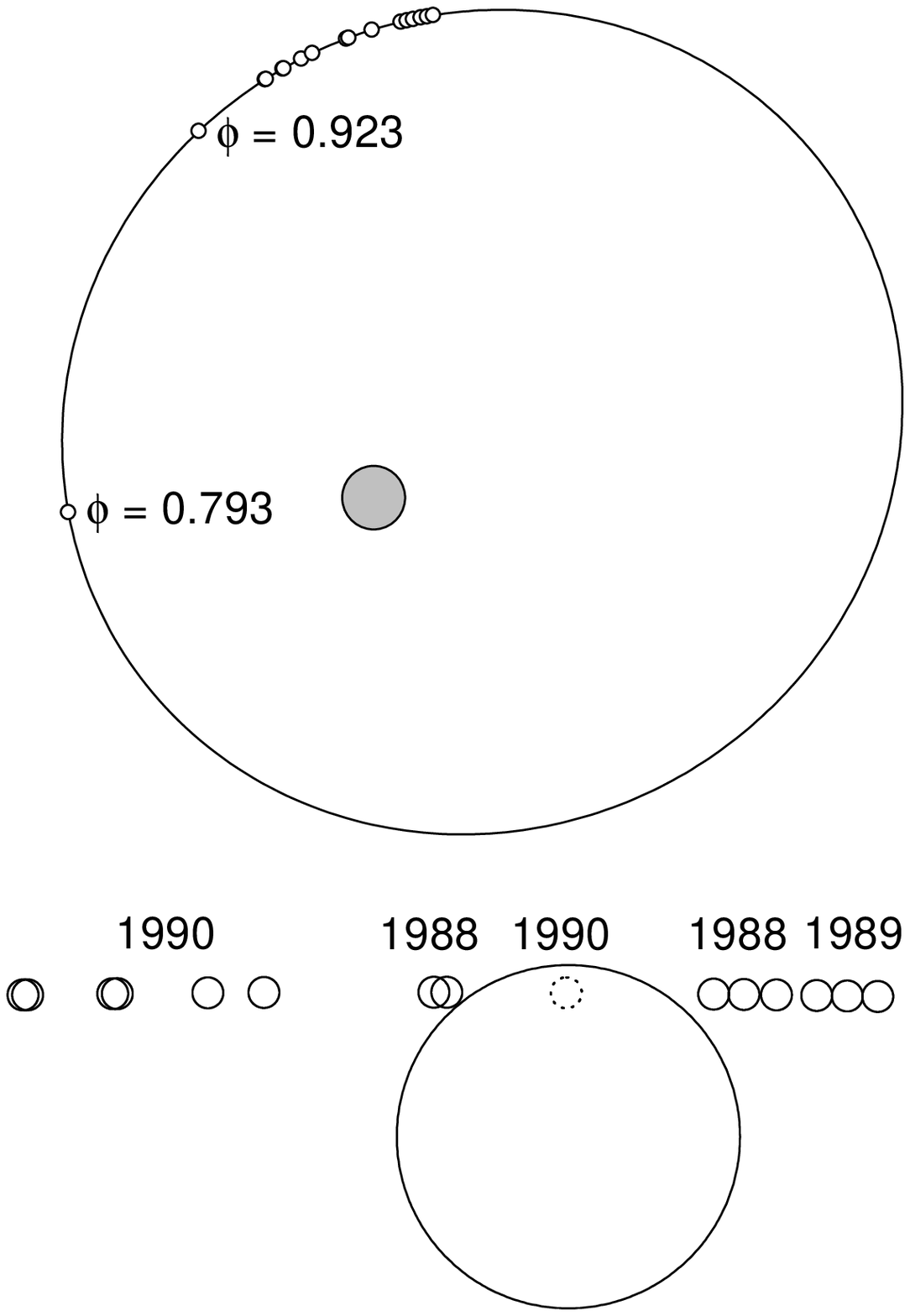}}
\caption[]{Relative orbit and binary geometry for all phases 
with detectable \ion{Si}{iv} and \ion{C}{iv} lines}
\label{fig3}
\end{figure}

Griffin et al.\ (1995) have redetermined the orbital and stellar parameters of HR~6902 
by means of photometric observations during the eclipses of 1987 and 1989 (Table~\ref{tbl-2}). 
The relative orbit and the eclipse geometry are shown in Fig.~\ref{fig3}. It should be noted that
the stellar parameters of giants belonging to $\zeta$~Aur systems can be much better determined
than for single stars. A recent evolutionary analysis suggests that the bright giant primary 
of HR~6902 is already evolved into its blue loop phase (\cite{sch97}).

\section{Absorption line diagnostics}
\label{sec3}

A comparison of the total eclipse emission line spectrum with the out-of-eclipse 
observations shows immediately that the continuum flux is orders of magnitude 
larger than the scattered line flux. Hence, the circumstellar absorption profiles 
can be interpreted as pure absorption lines formed in the extended envelope of 
HR~6902. In the subsequent sections we present details of the analysis and 
interpretation of the spectra of HR~6902.

\subsection{Basic equations}
\label{sec31}

The line formation in HR~6902 is dominated by velocity fields on different scales.
To simplify the concept we introduce a line broadening velocity $v_{\rm b}$ 
containing both the thermal velocity $v_{\rm th}$ and a stochastic component 
$v_{\rm sto}$ associated with turbulence or other short-scale nonthermal motions. 
Assuming a Gaussian distribution of the stochastic velocity the spectroscopic line 
broadening parameter becomes $v_{\rm b}^2 = v_{\rm th}^2 + v_{\rm sto}^2$. The 
macroscopic velocity field is characterized by a spherically symmetric expansion 
described by an a priori unknown velocity law $v_{\rm wind}(r)$.

Throughout this paper we assume the line broadening velocity to be constant in the 
relevant parts of the envelope. It is convenient to define a dimensionless frequency 
variable $x = (\nu - \nu_0)/\Delta\nu_{\rm d}$, where $\nu_0$ is the line-center 
frequency and $\Delta\nu_{\rm d}=\nu_0 v_{\rm b}/{\rm c}$ the corresponding Doppler 
width. Macroscopic velocities are measured in the same units, i.e. 
$u(r)=v(r)/v_{\rm b}$. Regarding the B star as a point light-source of intensity 
$I_{\rm B}$ the emergent flux for a given impact parameter $p$ can be written
\begin{equation}
\label{eq-1} 
F_x (p)=I_{\rm B} \exp \left[-\tau_{x}(p)\right].
\end{equation}
The optical depth along a ray from the observer to the B star is given by
\begin{equation}
\label{eq-2}
\tau_{\rm x} (p) = k_{\rm 0} \int_{z_{min}}^{z_{max}} n(r) 
\Phi \left[x-\frac{z}{r}u(r)\right]\, \rm {d}z,
\end{equation}
where $\Phi(x)$ denotes the dimensionless profile function. The geometry is 
defined by the distance from the primary $r$ and the integration path 
$z = \sqrt{r^2-p^2}$. The boundaries $z_{min}$ and $z_{max}$ depend on the 
position of the B~star and the adopted outer radius of the envelope (chosen
as large that the results are unaffected by its actual value), respectively. 
The line opacity per absorber reads
\begin{equation}
\label{eq-3}
k_{\rm 0} = \frac{\pi \rm {e^{2}}}{\rm {m_{e}} c} \lambda_{0} 
f v^{-1}_{\rm sto},
\end{equation}
where $f$ is the oscillator strength and $\lambda_{0}$ the rest wavelength of 
the transition under consideration.

In the static case (i.e. without considering the expansion explicitly) 
Eq.~\ref{eq-2} degenerates to a simple relation between the optical depth 
and the column density $N$, viz
\begin{equation}
\label{eq-4}
\tau_{\rm x} (p) = k_{\rm 0} N \Phi (x). 
\end{equation}

Our analysis is carried out using a non-linear least square fit technique based 
on the Levenberg-Marquard method (e.g., \cite{pre92}) to derive the column density 
$N$ and the line broadening parameter $v_{\rm sto}$. The {\it NEWSIPS\/} error
estimate provides statistical errors of the deduced parameters and allows
to assess the goodness-of-fit. As an additional constraint 
the best-fit parameters are forced to be equal for all lines of the particular 
multiplet under consideration, i.e., we favor the strategy of resonance doublet 
modeling instead of analyzing individual absorption lines. Our fitting procedure 
allows to vary the continuum flux level in order to avoid the difficulties 
associated with a continuum definition by eye. Finally, the theoretical profile 
has to be convoluted with the {\it IUE\/} line-spread function. The 
instrumental profile is assumed to be well represented by a Gaussian with FWHM 
in accordance with the actual determination of the {\it NEWSIPS\/} resolution 
(\cite{gar97}). 

\subsection{Quasi-static approach}
\label{sec32}

As a starting point of our analysis we use a quasi-static approximation neglecting 
macroscopic velocity fields. The objective of this approach is to provide a 
provisional set of column densities and stochastic velocities. However, line 
broadening effects due to the wind outflow calls the static model into question. 
In Sect.~\ref{sec33} we will examine possible consequences for the derived parameters.

\begin{figure}
\resizebox{\hsize}{!}{\includegraphics{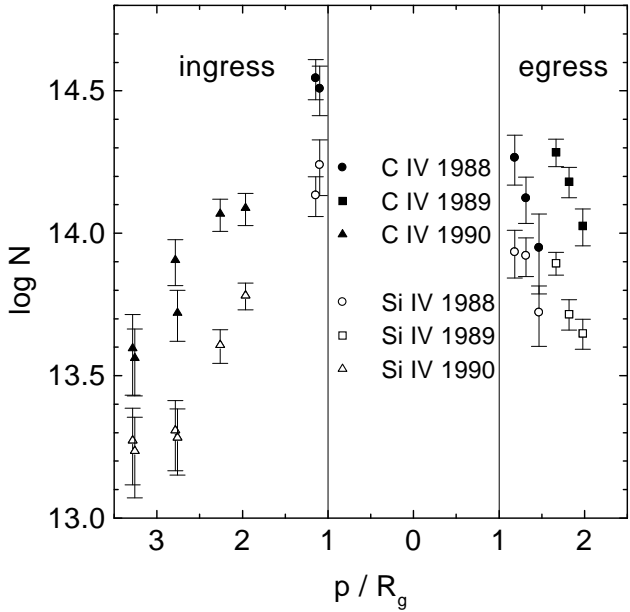}}
\caption[]{Column densities of $\ion{C}{iv}$ and $\ion{Si}{iv}$ obtained by the quasi-static approach}
\label{fig4}
\end{figure}

In the spectrum SWP 31325 (impact parameter $p = 5.2\; {\rm R_g}$) we cannot 
identify any signature of \ion{C}{iv} or \ion{Si}{iv} absorption lines. 
Instead, we use the error information of the {\it IUE\/} flux to determine 
upper limits of the column densities. The results of the
static analysis are summarized in Table~\ref{tbl-3}. 
Fig.~\ref{fig4} shows the run of column densities as a function of 
the impact parameter. As already noticed by Reimers et al.\ (1990b), the egress 
observations indicate significant differences from epoch to epoch. The obtained 
stochastic velocities show a considerable scatter with a weighted mean of 
$68 \pm 16\;{\rm km\, s}^{-1}$.

\begin{table*}
\caption[]{Column densities, stochastic velocities, and temperatures derived from
least square profile modeling.
The parameters are obtained using two distinct wind models. At each phase the 
first line displays the results of the quasi-static approach. The second line shows the effect
of the macroscopic velocity field ($\beta$~power-law with $v_{\infty} = 100\;\mathrm{km\, s^{-1}}$). 
Additional information can be found in the text}
\begin{flushleft}
\begin{tabular}{cccccc}
\hline
\noalign{\vspace {0.1cm}}
 Phase &$ \log N[\ion{Si}{iv}]$ & $\log N[\ion{C}{iv}]$ 
 & $v_\mathrm{sto} (\mathrm{km\,s^{-1}})$ & $T (10^3\,\mathrm{K})$
 & $\log N [\mathrm{H}]$ \\
\noalign{\vspace {0.1cm}}
\hline
\noalign{\vspace {0.1cm}}
$0.9227$ &$\le 12.95   $                  &$\le 13.30   $        &           --            &         --           &      --  \\
$0.9534$ &$13.27$ ($+0.11, -0.15$) &$13.60$ ($+0.12, -0.16$) &$101.9$ ($\pm$ 23.1) &$76.4$ ($\pm$5.0) &$18.08$ ($+0.18, -0.11$) \\
         &$13.25$ ($+0.10, -0.13$) &$13.59$ ($+0.10, -0.13$) &$62.7$ ($\pm$ 29.4) &$76.4$ ($\pm$5.3)  &$18.06$ ($+0.15, -0.12$)  \\
$0.9538$ &$13.24$ ($+0.12, -0.16$) &$13.56$ ($+0.10, -0.13$) &$64.0$ ($\pm$ 23.0) &$76.4$ ($\pm$4.5) &$18.04$ ($+0.18, -0.12$) \\
         &$13.23$ ($+0.11, -0.14$) &$13.59$ ($+0.10, -0.13$) &$65.5$ ($\pm$ 31.7) &$77.3$ ($\pm$5.6) &$18.04$ ($+0.17, -0.14$)  \\
$0.9612$ &$13.31$ ($+0.11, -0.14$) &$13.90$ ($+0.07, -0.09$) &$103.3$ ($\pm$ 16.2) &$82.2$ ($\pm$3.2) &$18.16$ ($+0.19, -0.12$)\\
         &$13.32$ ($+0.10, -0.13$) &$13.91$ ($+0.06, -0.07$) &$74.0$ ($\pm$ 20.3) &$82.2$ ($\pm$3.7)  &$18.17$ ($+0.16, -0.16$) \\
$0.9616$ &$13.28$ ($+0.10, -0.13$) &$13.72$ ($+0.08, -0.10$) &$58.5$ ($\pm$ 12.5) &$78.9$ ($\pm$3.2) &$18.10$ ($+0.15, -0.11$)\\
         &$13.30$ ($+0.10, -0.12$) &$13.77$ ($+0.08, -0.09$) &$48.8$ ($\pm$ 22.3) &$79.4$ ($\pm$4.1) &$18.13$ ($+0.16, -0.16$)\\
$0.9694$ &$13.61$ ($+0.05, -0.06$) &$14.07$ ($+0.05, -0.06$) &$64.2$ ($\pm$ 5.9) &$79.4$ ($\pm$1.9) &$18.43$ ($+0.08, -0.06$) \\
         &$13.59$ ($+0.05, -0.06$) &$14.02$ ($+0.05, -0.06$) &$42.1$ ($\pm$ 11.8)&$78.9$ ($\pm$2.4) &$18.40$ ($+0.08, -0.07$) \\
$0.9742$ &$13.78$ ($+0.04, -0.05$) &$14.09$ ($+0.05, -0.06$) &$67.7$ ($\pm$ 6.9) &$75.9$ ($\pm$1.8) &$18.59$ ($+0.05, -0.04$)\\
         &$13.74$ ($+0.04, -0.05$) &$14.03$ ($+0.05, -0.05$) &$32.3$ ($\pm$ 10.3)&$75.7$ ($\pm$2.3) &$18.56$ ($+0.05, -0.04$) \\
$0.9887$ &$14.13$ ($+0.06, -0.08$) &$14.54$ ($+0.07, -0.09$) &$71.9$ ($\pm$ 4.9) &$78.3$ ($\pm$2.4) &$18.95$ ($+0.09, -0.07$)\\
         &$14.11$ ($+0.07, -0.08$) &$14.55$ ($+0.07, -0.08$) &$46.1$ ($\pm$ 9.4)  &$78.9$ ($\pm$3.0)&$18.93$ ($+0.10, -0.09$) \\
$0.9898$ &$14.24$ ($+0.09, -0.11$) &$14.51$ ($+0.08, -0.10$) &$64.1$ ($\pm$ 4.9) &$75.0$ ($\pm$3.2) &$19.05$ ($+0.13, -0.10$)\\
         &$14.18$ ($+0.08, -0.10$) &$14.54$ ($+0.09, -0.11$) &$39.3$ ($\pm$ 11.4) &$77.3$ ($\pm$4.3)&$18.99$ ($+0.13, -0.10$)\\
$0.0122$ &$13.93$ ($+0.08, -0.09$) &$14.27$ ($+0.08, -0.10$) &$51.4$ ($\pm$ 5.0) &$76.6$ ($\pm$3.0)&$18.74$ ($+0.10, -0.07$)\\
         &$13.92$ ($+0.07, -0.08$) &$14.19$ ($+0.07, -0.08$) &$26.2$ ($\pm$ 12.9)&$75.0$ ($\pm$3.6) &$18.73$ ($+0.08, -0.05$)\\
$0.0147$ &$13.92$ ($+0.06, -0.07$) &$14.12$ ($+0.08, -0.09$) &$55.6$ ($\pm$ 5.3) &$73.3$ ($\pm$2.8)&$18.75$ ($+0.11, -0.09$)\\
         &$13.87$ ($+0.06, -0.07$) &$14.07$ ($+0.07, -0.08$) &$29.5$ ($\pm$ 13.6) &$73.4$ ($\pm$3.4)&$18.69$ ($+0.07, -0.05$)\\
$0.0175$ &$13.72$ ($+0.09, -0.12$) &$13.95$ ($+0.12, -0.16$) &$45.2$ ($\pm$ 12.4)&$74.1$ ($\pm$4.8) &$18.54$ ($+0.19, -0.13$)\\
         &$13.67$ ($+0.09, -0.11$) &$13.81$ ($+0.15, -0.23$) &$34.7$ ($\pm$ 24.1) &$72.0$ ($\pm$6.2) &$18.50$ ($+0.17, -0.12$)\\
$0.0209$ &$13.89$ ($+0.04, -0.04$) &$14.28$ ($+0.05, -0.05$) &$101.2$ ($\pm$ 6.7) &$78.0$ ($\pm$1.5) &$18.70$ ($+0.05, -0.05$)\\
         &$13.89$ ($+0.04, -0.04$) &$14.27$ ($+0.04, -0.05$) &$71.8$  ($\pm$ 9.5) &$77.8$ ($\pm$1.8) &$18.70$ ($+0.05, -0.04$)\\
$0.0235$ &$13.72$ ($+0.05, -0.06$) &$14.18$ ($+0.05, -0.06$) &$101.2$ ($\pm$ 15.0) &$79.4$ ($\pm$1.7)&$18.54$ ($+0.07, -0.06$)\\
         &$13.71$ ($+0.04, -0.05$) &$14.17$ ($+0.05, -0.05$) &$74.7$  ($\pm$ 17.5) &$79.3$ ($\pm$2.0) &$18.53$ ($+0.07, -0.06$)\\
$0.0261$ &$13.65$ ($+0.05, -0.06$) &$14.03$ ($+0.06, -0.07$) &$95.9$ ($\pm$ 9.0) &$77.6$ ($\pm$2.0) &$18.46$ ($+0.06, -0.05$)\\
         &$13.64$ ($+0.05, -0.05$) &$14.03$ ($+0.06, -0.06$) &$59.4$ ($\pm$ 15.4) &$77.8$ ($\pm$2.6) &$18.45$ ($+0.07, -0.06$)\\
\noalign{\vspace {0.1cm}}
\hline
\end{tabular}
\end{flushleft}
\label{tbl-3}
\end{table*}

Assuming the simplifying conditions of a low density plasma the \ion{C}{iv} and 
\ion{Si}{iv} density ratio can be used to derive an ionization temperature $T_\mathrm{ion}$. 
We solve the ionization balance using the radiative-collisional equilibrium program 
CLOUDY (version 9003; \cite{fer97}). The metal abundances are supposed to be solar 
($A_{\rm C}=3.6 \times 10^{-4}, A_{\rm Si}=3.6 \times 10^{-5}$ [\cite{and89}]). 
As an approximation we identify the ionization temperature with the kinetic temperature
$T_\mathrm{e}$ knowing that the ionization state may freeze in, so that we have
$T_\mathrm{e} \neq T_\mathrm{ion}$ in the outer wind.
It turns out that the electron temperature appears to be constant within the 
observed part of the envelope ($p = 1.1 - 3.3\; {\rm R_g}$) with a weighted mean 
of $T_{\rm e} = 78\,000 \pm 2000\;{\rm K}$. However, the ion ratio may be affected 
by different CNO abundances due to the first dredge up which can deplete the carbon abundance 
significantly (e.g., \cite{lam81}). For example, a carbon abundance 
reduced by a factor of 4 would lead to a temperature of $\sim 90\,000\;{\rm K}$.

\subsection{Density models}
\label{sec33}

With known electron temperature we are able to transform the ion density into 
total hydrogen densities. In the next step we try to match the derived hydrogen column 
densities employing different types of density laws. Each density law implies a physical 
model defined by a specific set of parameters. Again we implement a least square 
fit procedure using the Levenberg-Marquard method. 

\subsubsection{Hydrostatic model}
\label{sec331}

As a first attempt we use the hydrostatic equation for an isothermal plasma
\begin{equation}
\label{eq-5}
n_{\rm H}(r) = n_{\rm 0}\exp\left(-\frac{r_0}{r}\,\frac{r-r_0}{H}\right),
\end{equation}
where $r_0$ defines the bottom of the envelope with the corresponding hydrogen 
density $n_{\rm 0}$. $H$ denotes the density scale height. This approach 
is motivated by the fact that at least the subsonic region  will be well 
represented by the 
hydrostatic density law. In order to fit the empirical column densities we adjust 
the density parameter $n_{\rm 0}$ and the scale height $H$. Identifying $r_0$ 
with the giant radius the least square procedure yields $n_{0} = 5.0 \pm 0.92 
\times 10^{6}\; \rm{cm^{-3}}$ and $H = 3.4 \pm 0.14\times 10^{11}\; {\rm cm}$.

\begin{figure}
\resizebox{\hsize}{!}{\includegraphics{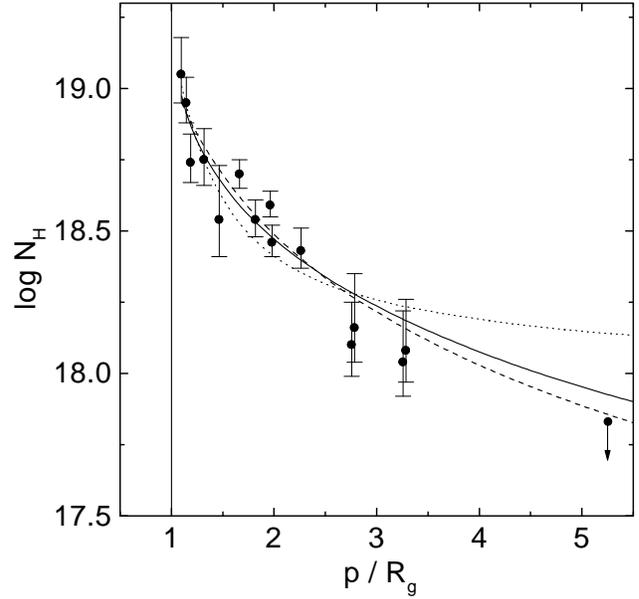}}
\caption[]{Comparison of best-fit density models 
with empirical hydrogen column densities (dotted: hydrostatic approach; solid: 
$\beta$~power-law; dashed: $P \propto \rho$ model)}
\label{fig5}
\end{figure}

The pressure may be composed of two parts, the thermal gas component and 
an effective turbulent pressure leading to the total dynamical pressure
\begin{equation}
\label{eq-6} 
P_{\rm tot}=\rho\left(\frac{{\rm k}T}{\mu m_{\rm H}}+\frac{v_{\rm turb}^{2}}{2}\right), 
\end{equation}
where $\mu$ is the mean molecular weight measured in units of the hydrogen mass.
The velocity $v_{\rm turb}$ denotes an isotropic turbulent motion 
with a spatial scale smaller than the pressure scale height.
Assuming the ionization equilibrium to be described by the physics of a low 
density plasma with $T_{\rm e}=78\,000\;{\rm K}$ we obtain $\mu=0.64$ [for solar 
composition according to Anders \& Grevesse (1989)]. If we identify the stochastic 
velocity of $68\; {\rm km\, s^{-1}}$ with the turbulent velocity we obtain a scale 
height of $3.4 \times 10^{11}$ cm, in agreement with the best-fit parameter 
of the hydrostatic model. However, at the impact parameter of $p = 5.2\; {\rm R_g}$ 
the theoretical column density exceeds the observed upper limit by a factor of 3 
(see Fig.~\ref{fig5}). As expected the hydrostatic equation provides an adequate description 
of the inner shell ($p < 3\; {\rm R_g}$), but overestimates the densities in the 
outer envelope considerably. 

\subsubsection{Kinematic model: $\beta$~power-law}
\label{sec332}

A more realistic model of the large-scale density distribution requires an 
appropriate kinematic approach. We follow two different strategies to introduce 
a velocity law: an analytical $\beta$~power-law approach traditionally used in 
$\zeta$ Aurigae studies, and a hydrodynamic description of an isothermal, 
isoturbulent wind (where $P/\rho = \mathrm{const}$). For the sake of simplicity we 
adopt a steady, spherically symmetric outflow. We are aware of the fact that 
these assumptions may be a point of criticism, since the observations suggest 
a considerable variability. However, the amplitude of these variations are small 
and justify a kind of "average" description.

The first kinematic model uses a velocity law proven to be adequate for different 
classes of stellar winds including the envelopes of evolved late-type stars.
The analyses of several $\zeta$~Aur binaries suggest a velocity field of the form
\begin{equation}
\label{eq-7}
v(r) = v_{\infty} \left(1 - \frac{\rm R_g}{r}\right)^\beta,
\end{equation}
where $v_{\infty}$ is the terminal wind velocity and $\beta$ the acceleration 
parameter which defines the steepness of the spatial velocity gradient associated 
with the expansion. The wind velocity $v(r)$ increases monotonically outward and 
approaches the terminal velocity at large distances from the giant. The number 
density is assumed to vary in accordance to the equation of continuity. 
The hydrogen density becomes
\begin{equation}
\label{eq-8}
n_{\rm H}(r) = n_{\beta} \left(\frac{\rm R_g}{r}\right)^{2} 
\left(\frac{r}{r-{\rm R_g}}\right)^\beta,
\end{equation}
where the density parameter $n_{\beta}$ is related to the mass-loss rate and 
terminal wind velocity:
\begin{equation}
\label{eq-9}
n_{\beta} = \frac{\dot{M}}{4\pi v_{\infty} {\rm R^2_g}} \frac{X_{\rm H}}{m_{\rm H}}.
\end{equation}
The hydrogen abundance by weight $X_{\rm H}$ is fixed at $0.70$ assuming solar 
chemical composition. The least square procedure yields a set of best-fit parameters 
$n_{\beta} = 6.6 \pm 0.7 \times 10^{5}$ $\rm{cm^{-3}}$ and $\beta = 0.60 \pm 0.14$. 

The small value of $\beta$ indicates a rapid acceleration to $\sim 0.5\, v_{\infty}$ 
within $1.5\; {\rm R_g}$. We note that the wind models of "classical" $\zeta$~Aurigae
systems ($\zeta$~Aur, 31 Cyg , 32 Cyg) indicate a gradual acceleration with $\beta$ 
in the range between 2.5 and 3.5 (e.g., \cite{sch85}). In the case of HR 6902 the 
terminal velocity eludes a spectroscopic determination, since we can only observe 
the tangential velocity component. The poor quality of the spectra does not
allow the use of alternative velocity indicators like \ion{Mg}{ii} absorption. 
As a consequence the mass-loss information
is restricted to the ratio $\frac{\dot{M}}{v_{\infty}}$. The winds from hybrid stars 
show high terminal velocities (typically 50 to $150\;{\rm km\,s^{-1}}$) as measured 
by the circumstellar \ion{Mg}{ii} absorption (e.g., \cite{dup87}). Furthermore, 
the flow velocities of luminous cool stars are always less than the escape speed 
from the photosphere (for HR~6902 we find $v_{\rm esc} = 211\;{\rm km\,s^{-1}}$).
So we expect $v_{\infty}$ to lie in the range $50 - 200\;{\rm km\,s^{-1}}$. The 
corresponding mass-loss rate is $8 \times 10^{-12} - 3.4 
\times 10^{-11}\;{\rm M_{\odot}\,yr^{-1}}$ with a statistical error of $\sim 10\%$. 

\subsubsection{Kinematic model: $P \propto \rho$ wind}
\label{sec333}
Finally, we try to attack the mass motion using the equations of gas dynamics. 
Conservation of momentum gives the one-dimensional Euler equation
\begin{equation}
\label{eq-10}
v \frac{{\rm d}v}{{\rm d}r}=-\frac{1}{\rho} \frac{{\rm d}P_{\rm eff}}{{\rm d}r}
-\frac{{\rm G}{\rm M_g}}{r^2}+f(r),
\end{equation} 
where $f(r)$ denotes a potential extra force per unit-mass. In the present study we
set $f=0$ and introduce an effective pressure term containing an unspecified
nonthermal contribution assuming a proportionality of the form 
$P_{\rm eff} \propto \rho$.  This simplification leads to a Parker-type wind. The 
Euler equation can be integrated immediately and yields a solution with a positive 
velocity gradient at all distances (e.g., \cite{mih78}):
\begin{equation}
\label{eq-11}
\frac{v}{v_{\rm c}} \exp \left[-\frac{1}{2}\left(\frac{v}{v_{\rm c}}\right)^2\right]
= \left(\frac{r_{\rm c}}{r}\right)^2 \exp \left(-\frac{2r_{\rm c}}{r} + \frac{3}{2}\right).
\end{equation}
The transcendental equation can be solved numerically to give $v(r)/v_{\rm c}$.
The critical radius is defined by the singularity of the Euler equation:
\begin{equation}
\label{eq-12}
r_{\rm c}=\frac{{\rm G}{\rm M_g}}{2v^{2}_{\rm  c}}.
\end{equation}
The hydrogen density may be expressed by an equation of the form
\begin{equation}
\label{eq-13}
n_{\rm H}(r) = n_{\rm c}\left(\frac{\rm R_g}{r}\right)^{2} \frac{v_{\rm c}}{v(r)},
\end{equation}
where the density parameter $n_{\rm c}$ is related to the mass-loss rate by
\begin{equation}
\label{eq-14}
n_{\rm c} = \frac{\dot{M}}{4\pi v_{\rm c} {\rm R}^{2}_{\rm g}} \frac{X_{\rm H}}{m_{\rm H}}.
\end{equation}
This formalism suggests varying the density parameter and the critical velocity to model
$n_\mathrm{H}(r)$.
The critical velocity specifies the singularity of the momentum equation
and thus determines the slope of the resulting density distribution.
We note that there is no need to take explicitly care of specific pressure constituents.
Our least square fit procedure yields $n_\mathrm{c}=1.5\pm 0.3\times 10^{6}\;
{\rm cm^{-3}}$ and $v_\mathrm{c}=89.5\pm 
13.7\;{\rm km\,s^{-1}}$. The density parameter leads to a mass-loss rate of
$\dot{M}=3.4\pm 1.2\times 10^{-11}\; \rm M_{\odot}\,yr^{-1}$.

Using Eq.~\ref{eq-12} the critical radius becomes $\sim 1.4\;{\rm R_g}$, 
indicating a high velocity close to the surface of the giant. Fig.~\ref{fig5} demonstrates
that both wind models lead to similar density laws. Indeed, the approximate relationship 
between $v_{\rm c}$ and $\beta$ derived by Harper et al.\ [1995, Eq. (10)] yields 
an acceleration parameter of 0.54 quite similar to the result of our beta-law 
analysis. This finding may justify the $\beta$ power-law approach in other red 
giant winds, where we have an unknown hydrodynamic situation.

Assuming the effective pressure to be supplied by both gas and turbulence pressure the
critical velocity is given by
\begin{equation}
\label{eq-15}
v_{\rm c}=\sqrt{\frac{{\rm k}T}{\mu m_{\rm H}} + \frac {v^{2}_{\rm  turb}}{2}}.
\end{equation}
The best-fit model with $v_{\rm c}=89.5\;{\rm km\,s^{-1}}$ yields a turbulence of
$v_{\rm turb}=119\;{\rm km\, s^{-1}}$ 
which exceeds the observed stochastic velocity by a factor of 1.8. This discrepancy 
may become even worse when we allow for macroscopic line broadening. Either the 
identification of turbulent and stochastic velocity is completely inappropriate, 
or the momentum Eq.~\ref{eq-10} contains unknown external forces or pressure 
terms. Furthermore, the relation between the observed 
line broadening velocity and the turbulent velocity is not clear and depends on 
the driving mechanism(s), the geometry, and the statistics of the absorbing 
structures.

In the preceding analysis we have ignored line broadening due to the wind 
velocity. In order to examine the effect on the derived parameters we repeat 
the absorption line analysis with the complete optical depth quadrature according 
to Eq.~\ref{eq-2}. The wind outflow is assumed to be well described by the $\beta$ 
power-law. Again we consider terminal velocities ranging from 50 to 
$200\;{\rm km\, s^{-1}}$. It turns out that high velocity models with 
$v_{\infty} \gtrsim 150\;{\rm km\, s^{-1}}$ cannot reproduce the observed 
profile shapes adequately. The deduced stochastic component of the line broadening
lies in the interval $20 - 100\;{\rm km\, s^{-1}}$ showing partly a large 
standard error. The range of stochastic velocities reflects both noisy 
{\it IUE\/} spectra and true irregularities of the local velocity field. 
As an example we present in Table~\ref{tbl-3} the results for $v_{\infty}=100\;{\rm km\, 
s^{-1}}$ which represents the most probable fit model. Though this model yields 
considerably reduced stochastic velocities with a weighted mean (i.e., averaged 
over all heights) of $47 \pm 16\;{\rm {km\, s}^{-1}}$, the derived column densities 
are fairly close to the quasi-static model. This finding even holds for larger 
wind speeds and we conclude that the quasi-static analysis has produced column 
densities with a reasonable degree of confidence.

\section{Emission line diagnostics}
\label{sec4}

Reimers et al.\ (1990a) have shown that the \ion{C}{iv} and \ion{Si}{iv} emission 
lines are formed by resonance scattering of B star photons in the envelope of 
the primary. Profile modeling requires the solution of a two-dimensional radiative 
transfer problem. The difficulty arises due to the fact that the light source 
(i.e. the B star) is displaced from the center of wind symmetry. It is well-known that 
the scattering lines are formed in a large volume compared to the scale of the 
binary orbit (\cite{baa96}). Hence, the emission line diagnostics allows to study 
the large-scale properties of the circumstellar envelope. 
 
For the 2-D radiative transfer calculations we employ the SEI (Sobolev with exact 
integration) method, i.e., the source function is approximated using the escape 
probability formalism followed by an exact formal solution using short characteristics. 
This procedure is simple in concept and can be applied to very extended geometrical 
grids. In normal $\zeta$~Aur systems the Sobolev approach leads to erroneous 
results, since the wind velocities are typically only a few times the stochastic 
velocities. However, a comparison with a more refined 2-D radiative transfer code 
(\cite{baa89}, 1990) demonstrates the reliability of the Sobolev solution in the 
case of HR~6902. The reason is the low mass-loss rate ($\dot{M}\approx 10^{-11}
\;{\rm M_{\odot}\,yr^{-1}}$) compared to the "classical" $\zeta$~Aur systems 
($\dot{M}\approx 10^{-9}-10^{-8}\;{\rm M_{\odot}\,yr^{-1}}$). As a consequence the 
line opacities are small and multiple line photon scattering is of minor importance. 

\subsection{Sobolev approximation}
\label{sec41}

In the framework of the first order escape probability theory the scattering integral 
is given by (e.g., \cite{mih78})
\begin{equation}
\label{eq-16}
\left<{J}\right> = [1-\beta({\bf r})]S({\bf r})+\beta_{\rm c}({\bf r})I_{\rm c}, 
\end{equation}
where $I_{\rm c}$ denotes a constant core intensity representing the B star. 
The escape probabilities are given by
\begin{equation}
\label{eq-17}
\beta({\bf r}) = \frac{1}{4\pi} \int_{4\pi} \frac 
{1-\exp\left(-\tau_{\rm s}\right)}{\tau_{\rm s}} \,{\rm d} \Omega, 
\end{equation}
and
\begin{equation}
\label{eq-18}
\beta_{\rm c}({\bf r}) = \frac{1}{4\pi} \int_{\Omega_{\rm c}} \frac 
{1-\exp\left(-\tau_{\rm s}\right)}{\tau_{\rm s}} \,{\rm d} \Omega, 
\end{equation}
with the Sobolev optical thickness  $\tau_{\rm s} = k_{0}n({\bf r}) L({\bf r},{\bf n})$. 
The so-called Sobolev length in the direction ${\bf n}$ may be defined in tensor 
notation:
\begin{equation}
\label{eq-19}
L({\bf r, n}) = \left|\sum\limits_{ij} n_{i} n_{j}\frac{\partial u_{i}}
{\partial r_{j}}\right|^{-1}. 
\end{equation}
The quantity $\beta({\bf r})$ denotes the probability that a photon emitted at 
the point $\bf {r}$ escapes from the outer boundary and $\beta_{\rm c}({\bf r})$ 
symbolizes the probability that a photon emitted by the core (i.e. the B star) 
penetrates to the point ${\bf r}$. In the case of pure resonance scattering the 
line source function is simply given by
\begin{equation}
\label{eq-20}
S({\bf r}) = \frac{\beta_{\rm c}({\bf r})I_{\rm c}}{\beta({\bf r})}. 
\end{equation}
Detailed information can be found in Hempe (1982) who applied the Sobolev formalism 
to the specific conditions of the binary configuration. With known source function 
on a two-dimensional grid, we can calculate the emergent radiation field. The 
intensities are supplied by the formal solution of the transfer equation and can be
integrated along a set of parallel rays through the envelope. In a subsequent 
step the integration over all solid angles and normalization yields the residual 
flux $F_{\rm x}/F_{\rm c}$. For a detailed description we refer to Baade (1990) 
and Kirsch \& Baade (1994). 

\subsection{Application to HR 6902}
\label{sec42}

To calculate synthetic emission lines we extrapolate the wind models derived in 
Sect.~\ref{sec3} to the whole envelope assuming the temperature and thus the ionization 
state to remain fixed. Though this simplistic approach implies spurious energy 
sources to maintain the temperature it may be sufficient to explain qualitatively 
the observed flux.

In the first step we use the \ion{C}{iv} and \ion{Si}{iv} line ratio observed 
during total eclipse to derive the ionization temperature. 
Unfortunately, the clearly visible \ion{C}{ii} line eludes the diagnostic procedure, 
since the photospheric background profile is not known precisely. Furthermore, 
the loci of line formation may be quite different. We calculate a grid of 
theoretical flux ratios in the temperature range between 60\,000 and $90\,000\;{\rm K}$. 
The ionization problem is solved using the CLOUDY code as described in Sect.~\ref{sec32}. 
To consider the influence of different wind velocities we have a closer look at four 
different models defined either by the $\beta$~power-law or the $P \propto \rho$ wind 
according to Sect.~\ref{sec33}.

\begin{figure}
\resizebox{\hsize}{!}{\includegraphics{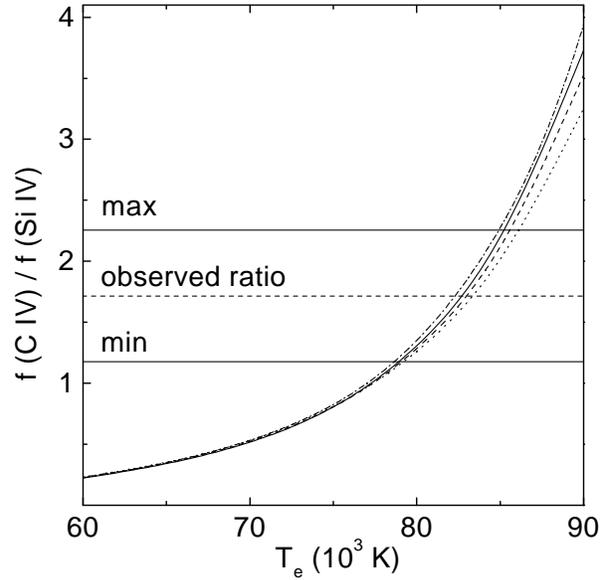}}
\caption[]{\ion{C}{iv}$/$\ion{Si}{iv} line flux ratio vs.\ electron temperature.
The observed ratio is obtained from the total eclipse spectrum SWP 39795.
The theoretical curves are presented for the $\beta$~power-law approach with different 
parameter sets [$\dot{M}({\rm M_{\odot}\,yr^{-1}}), v_{\infty}(\rm {km\, s^{-1}})$]:
Solid [$8\times 10^{-12}, 50$], dashed [$1.7\times 10^{-11}, 100$], dotted 
[$3.4\times 10^{-11}, 200$]. The $P \propto \rho$ model is indicated by a dashed-dotted
curve}
\label{fig6}
\end{figure}

As shown in Fig.~\ref{fig6} the theoretical flux ratios depend only slightly on the specific 
wind model. The observed line ratio suggests an electron temperature of $T_{\rm e} = 
82\,500 \pm 3000\;{\rm K}$ consistent with the mean temperature obtained on the basis 
of the absorption line analysis. We use the derived electron temperature to calculate 
integrated line fluxes of the \ion{C}{iv} and \ion{Si}{iv} emission lines. The observed 
values (Table~\ref{tbl-4}) can be reconstructed using the parameters $n_{\beta} = 2.9 
\times 10^{5}$ $\rm{cm^{-3}}$ for the $\beta$~power-law and $n_{\rm c} = 8.3 \times 
10^{5}$ $\rm{cm^{-3}}$ for the $P \propto \rho$ wind model. In both cases the density 
parameter differs only by a factor of $\sim 2$ from the results of the absorption line 
analysis. The discrepancy may be a result of the inadmissible extrapolation of the 
isothermal density distribution. Indeed, we would expect a decreasing temperature 
at large distances from the giant causing a reduced absorber density. 

\begin{table}
\caption[]{Observed emission line fluxes}
\begin{flushleft}
\begin{tabular}{lcc}
\hline
\noalign{\vspace {0.1cm}}
 Ion  & $\lambda $& f \\
  & (\AA)  &$ (10^{-14}\,\mathrm{erg\,cm^{-2}\,s^{-1}})$\\
\noalign{\vspace {0.1cm}}
\hline
\noalign{\vspace {0.1cm}}
$\ion{Si}{iv}$ & 1395/1408 & $14.2 \pm 3$\\
$\ion{C}{iv}$  & 1548/1550 & $24.7 \pm 4$\\
\noalign{\vspace {0.1cm}}
\hline
\end{tabular}
\end{flushleft}
\label{tbl-4}
\end{table}

\begin{figure}
\resizebox{\hsize}{!}{\includegraphics{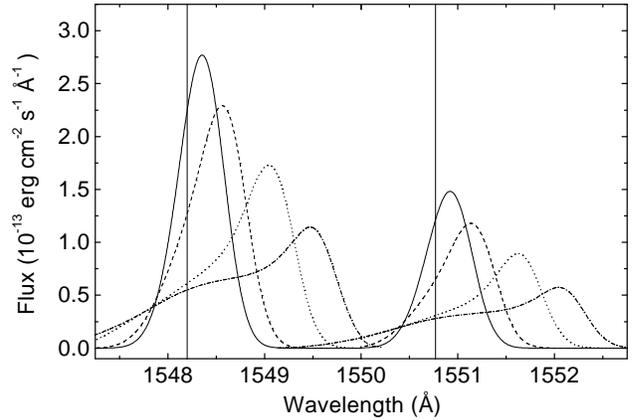}}
\caption[]{Theoretical line fluxes of the \ion{C}{iv} resonance 
doublet during total eclipse ($\phi = 1.000$). The line-center wavelengths are indicated 
by vertical lines. We present emergent profiles for the same wind models as shown in 
Fig.~\ref{fig6}}
\label{fig7}
\end{figure}

The presence of a macroscopic velocity field will influence the profile shapes
and cause wavelength shifts of the emission lines. However, the low resolution 
{\it IUE\/} spectra ($R \approx 6\;\mathrm{\AA}$) does not allow to examine these 
effects more closely. To demonstrate the expected situation we present in 
Fig.~\ref{fig7} a selection of theoretical flux profiles. It is obvious that the 
asymptotic flow velocity determines the character of the line profiles. In 
combination with a semi-empirical density model the line positions and profile 
shapes give reliable information about the global wind structure. This allows 
us to predict that high-resolution UV observations will unravel the kinematic 
nature of the circumstellar envelope of HR~6902.

\section{Summary and discussion}
\label{sec5}

\begin{figure}
\resizebox{\hsize}{!}{\includegraphics{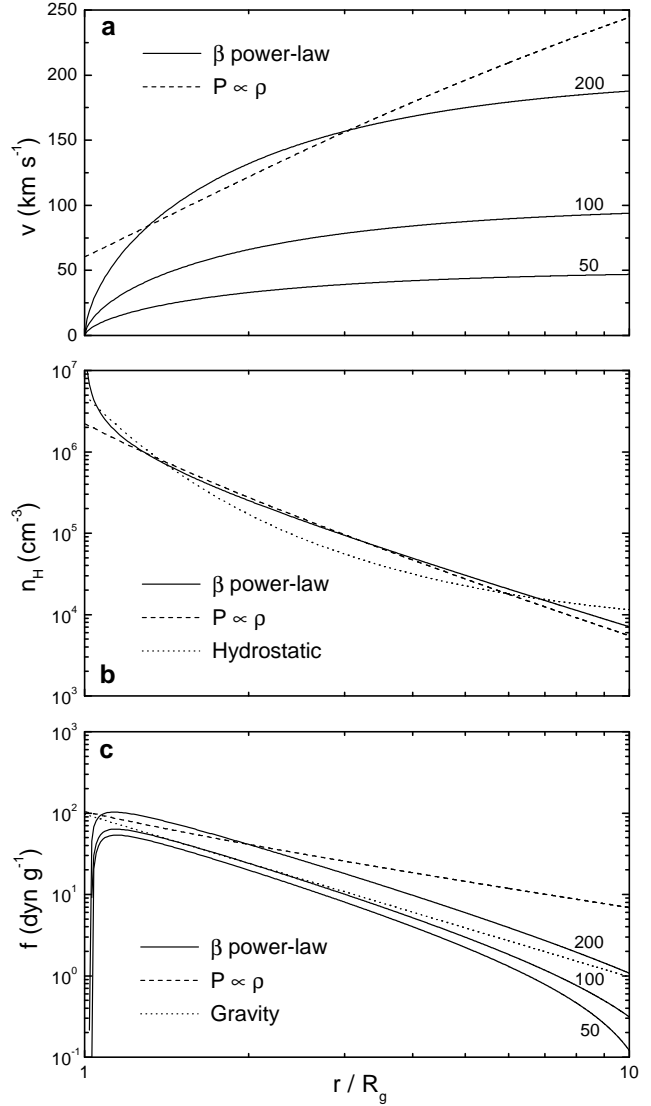}}
\caption[]{Velocity (a), density structure (b), and nonthermal momentum deposition (c) plotted 
as a function of radial distance. The $\beta$~power-law models are labelled with the corresponding 
terminal velocity}
\label{fig8}
\end{figure}

In the present paper we have analyzed {\it IUE\/} observations of the $\zeta$~Aurigae 
type binary system HR~6902. The hot secondary serves as a convenient probe for 
the extended envelope of the bright giant. By means of a least square profile-fit 
procedure we have deduced \ion{C}{iv} and \ion{Si}{iv} column densities and stochastic 
velocities. The density distribution of the inner envelope ($ \le 3\; {\rm R_{g}}$) 
can be described by a quasi hydrostatic model. We have shown that pressure support by 
turbulence is sufficient to explain the empirical scale height. However, in order 
to match the empirical column densities adequately we had to introduce an 
appropriate outflow model. We have demonstrated that both the common $\beta$~power-law 
and a $P \propto \rho$ wind yield appropriate density distributions and explain the absence 
of \ion{C}{iv} and \ion{Si}{iv} absorption at larger distances from the giant. Fig.~\ref{fig8} 
demonstrates that the hydrostatic solution as well as the different outflow 
models lead to similar density laws. Despite this similarity the hydrodynamic implications 
are completely different. The required driving force per unit mass $f(r)$ should be compared 
to theoretical models. 

We have derived mass-loss rates in the range $0.8 - 3.4\times 10^{-11}\; {\rm M}_{\odot}\,
{\rm yr^{-1}}$. This result indicates that the mass-loss rate of HR~6902 scales roughly 
with the solar mass flux, i.e., $\dot{M}\approx\ {\rm \dot{M}_{\odot}
(R_{g}/R_{\odot})^{2}}$. Finally, we have demonstrated that the emission line spectrum 
obtained during total eclipse confirms essentially the wind models derived from the 
absorption line analysis. This finding supports the picture of a steady and spherically
symmetric flow without excessive deviations from the mean wind model.

It is shown that HR~6902 possesses warm ($T_{\rm {e}}\approx 80\,000\; {\rm K}$) 
circumstellar material, observed in a range of impact parameters between 1.1 
and $3.3\; {\rm R_g}$. Recently, the innermost region immediately above the 
photosphere ($p \le 1.1\; {\rm R_{g}}$) has been examined by Schr\"oder, Marshall, 
\& Griffin (1996) using optical spectra. They derived a temperature of 
$T_{\rm {e}}\approx 5000\;{\rm K}$ and a turbulent velocity of $v_{\rm turb} = 
15\;{\rm km\, s^{-1}}$. However, in order to fit the \ion{Ca}{ii} K-line wing they 
postulated a second absorption component with a considerably larger turbulent 
velocity of $\sim 45\;{\rm km\, s^{-1}}$ quite similar to the stochastic velocities
derived in the present work. Combining these results we suspect a dramatic change 
of the physical conditions in the giant's atmosphere at $p \approx 1.1\; 
{\rm R_g}$ associated with a steep rise of the temperature from about 5000 to 
$80\,000\;{\rm K}$.

The large nonthermal line broadening parameter and its relative constancy (apart 
from local fluctuations) may be a characteristic property of the wind of HR~6902. 
The simplistic identification of turbulence and the observed stochastic velocity 
cannot explain the large pressure required to explain the density distribution 
by a $P \propto \rho$ wind. However, the true relation between the observed line width 
and turbulent motions associated with wave-driven winds depends on the specific
driving mechanism, geometrical details, and the randomness of the absorbing structures 
(e.g., \cite{jud92}).

The wind models suggest a rapid wind acceleration within $\sim 0.5\; {\rm R_g}$ above
the photosphere. This result is similar to the outflow characteristics of
hybrid stars [$\alpha$~Aqr (G2 Ib), $\beta$~Aqr (G0 Ib), $\theta$~Her (K1 IIa), 
$\iota$~Aur (K3 II), $\gamma$~Aql (K 3II), and $\alpha$~TrA (K3 II); \cite{bro86}; 
\cite{dup92}; \cite{har95}]. Furthermore, a recent analysis of {\it ORFEUS~II\/} 
spectra of the hybrid stars $\alpha$~TrA and $\alpha$~Aqr (\cite{dup98}) have 
revealed that the stellar winds are not confined to cool circumstellar material.
They pointed out that ions like \ion{Si}{iii} through \ion{O}{vi} indicate expanding 
material up to a temperature of $3\times 10^{5}\;{\rm K}$ with wind velocities 
from 100 to $200\;\rm {km\, s^{-1}}$. We argue that the wind properties of hybrid 
stars are comparable to the extended atmosphere of HR~6902.

With its location in the HR diagram HR~6902 represents the physically important 
connection between solar-like stars with hot coronae and late type supergiants 
with cool outer atmospheres. The binary technique provides spatially resolved 
information about the physical structure of the extended envelope and wind that 
are unobtainable by other means. With {\it IUE\/} we have gained a first insight, 
but additional high resolution UV observations are indispensable to construct a 
detailed model. With this unique opportunity HR~6902 may play a crucial role in 
the elucidation of atmospheric heating processes and mechanisms to drive winds of 
cool stars.
\\

\acknowledgements{}
This work is supported by the Verbundforschung of the Bundesministerium f\"ur 
Bildung, Wissenschaft, Forschung und Technologie under Grant No. 50 OR 96016. 
This research has made use of IUEFA data obtained through the ESA {\it IUE\/} 
data server. The authors would like to thank Graham Harper for discussions and
useful suggestions. Gary Ferland is kindly acknowledged for providing his ionization 
code CLOUDY. We made use of the ViziR search engine, a joint effort of CDS 
(Strasbourg, France) and ESA-ESRIN.

\end{document}